\documentclass[amsmath,amssymb,twocolumn,aps,pra]{revtex4-2}

\usepackage{times,color}

\usepackage{float}
\usepackage[table]{xcolor}
\usepackage{pgfplots}
\usepackage{graphicx}
\usepackage{dcolumn}
\usepackage{amsthm}
\usepackage{bm}
\usepackage{multirow}
\usepackage{algcompatible}
\usepackage{algorithm}

\newcommand{\nix}[1]{}
\usepackage{paralist}
\newtheorem{theorem}{Theorem}
 
\newtheorem{proposition}{Proposition}
\newtheorem{lemma}[theorem]{Lemma}

\newtheorem{remark}[theorem]{Remark}
\usepackage{subcaption}
\usepackage{standalone}
\usepackage{stackengine}
\usepackage{caption}
\usepackage{subcaption}

\usepackage[hidelinks,colorlinks=true,urlcolor= blue,linkcolor = blue,citecolor = red]{hyperref}

\begin{document}

\title{Decoding Topological Subsystem Color Codes Over the Erasure Channel using Gauge Fixing}

\author{Hiteshvi Manish Solanki}
\affiliation{%
 Department of Electrical Engineering, Indian Institute of Technology Madras, Chennai 600 036, India
}%
\author{Pradeep Kiran Sarvepalli}%
\affiliation{%
 Department of Electrical Engineering, Indian Institute of Technology Madras, Chennai 600 036, India
}%

\date{\today}

\begin{abstract}        
Topological subsystem color codes (TSCCs) are an important class of topological subsystem codes  that allow for syndrome measurement with only 2-body measurements. It is expected that such low complexity measurements can help in fault tolerance.  While TSCCs have been studied over depolarizing noise model, their performance over the erasure channel has not been studied as much. Recently, we proposed erasure decoders for TSCCs and reported a threshold of 9.7\%.  In this paper, we continue our study of TSCCS over the erasure channel. We propose two erasure decoders for topological subsystem color codes.  These decoders use the technique of gauge fixing where some of the gauge operators of the subsystem code are promoted to stabilizers. We perform gauge fixing using 4-body and 8-body gauge operators.  With partial gauge fixing we obtained a threshold of 17.7\% on  a TSCC derived from the square octagon lattice.  Using an order maximal gauge fixing decoder we were able to improve the threshold to  44\%. The previously known decoder for TSCC over the erasure channel had a threshold of 9.7\%.  
We also study the correctability of erasures on the subsystem codes.  
\end{abstract} 

\maketitle

\section{Introduction}
Subsystem codes are an important generalization of stabilizer codes \cite{bacon2006operator,kribs05,poulin2005stabilizer}. 
One of the motivations for this generalization was to simplify the error recovery process
of quantum codes. 
Specifically, they can simplify the syndrome computation which involves the measurement of the stabilizer generators.
These codes are defined by a (non-Abelian) subgroup of Pauli group called the gauge group. 
Of particular interest are the codes with local gauge group.
Topological subsystem color codes are a class of subsystem codes where generators of the gauge group as well 
the stabilizer of the code are  local.

The advantages of topological subsystem codes (TSCs) motivated many researchers to look at their performance under different noise models. 
Most of the studies on the  performance of TSCs \cite{bombin2012universal,suchara2011constructions, gayatri2018decoding,higgott2021subsystem} focused 
on the depolarizing channel. 
The error correcting capabilities of TSCCs for  handling leakage errors 
were explored in \cite{brown2019handling}. 
In contrast, the performance of TSCCs over the quantum erasure channel \cite{grassl1997codes}, which can be used to model qubit loss, has not been studied as much. 
Motivated by this gap, we  began the study of topological subsystem color codes over the erasure channel and reported some preliminary results in \cite{ITW_TSCC}.
Therein we proposed multiple two-stage erasure decoders for the TSCC derived from the square octagon lattice. 
Using a combination of various techniques like peeling  and clustering we were able to achieve a 
threshold of 9.7\% in \cite{ITW_TSCC} for the TSCC derived from the square octagon lattice. 
However, this threshold was not optimal and there was room for improvement. 

In this paper, we continue our study of TSCCs over the erasure channel and present new decoding algorithms that improve upon the decoders in \cite{ITW_TSCC}.
A driving force for this work is to improve the threshold of the TSCCs for erasure noise.
Topological subsystem color codes have fewer stabilizer generators than a comparable color code or a surface code. 
This will lead to lower thresholds \cite{bombin2012universal,suchara2011constructions,andrist12}. 
To address this we use the technique of gauge fixing, wherein we promote some of the gauge operators to checks. 
Gauge fixing  in effect leads to a larger stabilizer.
This technique of promoting some gauge operators to stabilizers was used for analyzing  the structure of subsystem codes in  \cite{bombin2012universal}, and for quantum computation in \cite{paetznick2013universal,bombin2015gauge}.
More recently, it was also employed  for decoding subsystem codes \cite{higgott2021subsystem}. 

An immediate question for the gauge fixing decoders is to determine which gauge operators to promote to stabilizers.
We propose two decoding algorithms each of which makes a different choice of the operators for gauge fixing. 
Both of them rely on the fact that a TSCC can be mapped to  color codes.
Furthermore, the  stabilizers of these color codes are in the gauge group of the TSCC.
We choose two different sets of these color code stabilizers for gauge fixing. 

Another question in the context of gauge fixing decoders is related to the sequence of measurement 
of the gauge operators. 
In our case, all the gauge operators that we promote to stabilizers anticommute with some 
2-body gauge operator of the TSCC. 
While we were able to obtain efficient decompositions of these additional stabilizers 
individually, these decompositions were not amenable to a joint measurement   using
2-body measurements. 
Therefore our decoding algorithms require direct measurement of these stabilizers. 
Then employing the mapping of the TSCC to color codes, we proceed to decode the TSCC. 

We briefly summarize our main contributions  below.
\begin{compactenum}[(i)]
    \item First, we propose a decoder that uses partial gauge fixing for TSCC. We obtain a threshold of $17.7\%$ for TSCC derived from the square octagon lattice. 
    
    \item We then propose an alternate ``order'' maximal gauge fixing decoder to improve the threshold. 
    This decoder leads to a threshold of threshold 44\%. 
    This improvement is attained at the expense of a slight increase in complexity.
    \item We study the correctability of erasures on a subsystem code. 
    Specifically, we provide a necessary and sufficient condition for an erasure pattern to be correctable on a subsystem code (without gauge fixing). We also study correctability of erasures under the order maximal gauge fixing decoder. 
\end{compactenum}

There are two related works in addition to our previous work \cite{ITW_TSCC}.
In \cite{brown2019handling}, the authors studied topological subsystem codes for correcting leakage errors, a more severe form of noise than erasure noise. 
While somewhat related, this noise model is different from ours and assumes that the locations of the erased qubits are
unknown. 
With respect to the technique of gauge fixing, the closest work is that of \cite{higgott2021subsystem}
where the authors  also use gauge fixing for decoding subsystem codes.
They studied subsystem surface codes and their generalizations over hyperbolic surfaces under depolarizing noise and 
biased noise models. 

We organize the paper as follows. In Section~\ref{sec:background} we review the background for our proposed decoders. 
In Section~\ref{sec:overview} we give an overview of the proposed decoding algorithms. 
Section~\ref{sec:synd_and_prepro} elaborates on syndrome measurement and preprocessing techniques.
In Section~\ref{sec:first_stage} we discuss the first stage for $X$ error correction and in Section~\ref{sec:second_stage} the second stage for $Z$ error correction. 
Then we study conditions for correctability of erasures on subsystem codes in  Section~\ref{sec:bound}.
We report the simulation results in Section~\ref{sec:simulation_results} and finally conclude with a brief summary in Section~\ref{sec:conclusion}.

In the previous version of this paper, we incorrectly claimed that the additional operators for gauge fixing can 
be jointly measured using 2-body and 4-body operators. This claim has been withdrawn in this version. We thank 
Oscar Higgot for pointing out the error.

\section{Background}\label{sec:background}
In this section, we briefly review some background material. 
We assume that the reader is familiar with stabilizer codes, see \cite{nielsen_chuang_2010,lidar_brun_2013} for an introduction.

\subsection{Subsystem codes} \label{subsec:subsystemCode}
We briefly review subsystem codes. 
For more details we refer the reader to \cite{lidar_brun_2013}.
Subsystem codes are obtained from stabilizer codes by not encoding information in some of the  logical qubits. 
These qubits are called the gauge qubits. 
Any error on the gauge qubits do not affect the codespace. 
An $[[n, k, r, d]]$ subsystem code encodes $k$ logical qubits and $r$ gauge qubits into $n$ qubits. 
It can detect errors upto $d-1$ qubits, where $d$ is the distance of the code. 
The distance also signifies the smallest weight of non-trivial logical operator. 

We define a subsystem code by a subgroup $\mathcal{G}\subset \mathcal{P}_n$, where $\mathcal{P}_n$ is the  Pauli group on $n$ qubits.
Elements of $\mathcal{G}$ are called the \textit{gauge operators}.  
Elements which generate the gauge group are called the \textit{gauge generators}. 
Recall that the centralizer of a subgroup $\mathcal{G}\subseteq \mathcal{P}_n$ is defined as
\begin{eqnarray}
C(\mathcal{G}) =  \{ g\in \mathcal{P}_n \mid  gh = hg \text { for all } h\in \mathcal{G} \}\label{eq:centralizer}
\end{eqnarray}
The \textit{stabilizer} $\mathcal{S}$ of the subsystem code is a subgroup of $\mathcal{G}$ such that $\langle iI,\mathcal{S} \rangle= \mathcal{G} \cap \mathcal{C(G)}$, where $\mathcal{C(G)}$ is the centralizer of $\mathcal{G}$.
Elements of $\mathcal{S}$ act trivially on the code space. 
It is convenient to ignore the phases and write  the stabilizer up to a phase as follows.
\begin{eqnarray}
\mathcal{S}= \mathcal{G} \cap C(\mathcal{G})\label{eq:stab-gauge-reln}
\end{eqnarray}
Since $\mathcal{S}\subseteq \mathcal{G}$, it follows that
\begin{eqnarray}
\mathcal{S}\subseteq \mathcal{G} \subseteq C(\mathcal{G})\subseteq C(\mathcal{S}).\label{eq:pure-logical-all-logical-reln}
\end{eqnarray}

For an $[[n, k, r, d]]$ subsystem code, ignoring phases, there are $2r+s$ independent generators for $\mathcal{G}$ and $s$ independent generators for $\mathcal{S}$, where  
 $n=k+r+s$, see \cite{poulin2005stabilizer}.
Operators in $C(\mathcal{G})\setminus \mathcal{G}$ are called bare logical operators while operators in 
$C(\mathcal{S})\setminus \mathcal{G} $, obtained by appending  gauge operators to the bare logical operators, are called dressed logical operators, see \cite{suchara2011constructions}. 

\subsection{Topological color codes}\label{subsec:TCC}
Topological color codes (TCCs) in 2D are defined on lattices where the vertices are trivalent and faces are 3-colorable \cite{bombin2006topological}. 
These lattices are also called 2-colexes. 
A quantum code can be defined on 2-colex by placing qubits on the vertices, and checks on faces. (We can shall refer to a 2-colex as a color code for this reason).
For any face $f$, a pair of stabilizers are defined as follows:
\begin{equation}
    B_f^X= \prod_{v \in f} X_v \hspace{4.2pt} \text{and}\hspace{4.2pt} B_f^Z= \prod_{v \in f} Z_v \label{eq:tcc-stabilizers}
\end{equation}
where $X$ and $ Z$ are the Pauli matrices. 
Let $F_\gamma$ be set of faces of $\gamma$ color, $\gamma \in \{r,g,b\}$.
These checks satisfy the following relations: 
\begin{subequations}\label{eq:CC_dependency}
        \begin{equation}
    \prod_{f \in F_r} B_{f}^Z=\prod_{f \in F_g} B_{f}^Z=\prod_{f \in F_b} B_{f}^Z \label{eq:tcc-z-stab-dep}
\end{equation}
\begin{equation}
    \prod_{f \in F_r} B_{f}^X=\prod_{f \in F_g} B_{f}^X=\prod_{f \in F_b} B_{f}^X \label{eq:tcc-x-stab-dep}
\end{equation}
\end{subequations}
This equality depicts that there are four dependencies among the stabilizer generators
given in Eq.~\eqref{eq:tcc-stabilizers}.
A color code defined on a 2-colex embedded on a surface of genus $g$ encodes $4g$ qubits.
For the rest of the paper we assume that the color code is defined on a torus which has genus one.
In this case the color code encodes 4 logical qubits. 

\begin{figure}[H]
    \centering
    \includegraphics[scale=0.8] {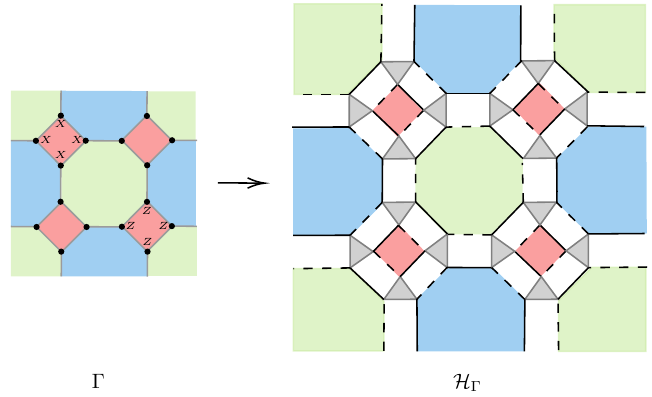}
    \caption{The figure on left shows the color code defined on 2-colex. Solid dots indicate the qubits. Every face has a $X$ type stabilizer and a $Z$ type stabilizer associated to it. A hypergraph $\mathcal{H}_{\Gamma}$ can constructed from this 2-colex. The hypergraph is depicted on the right. Shaded triangles correspond to rank-3 edges. A TSCC is defined on this hypergraph.  Qubits are located on the vertices of the hypergraph and its edges define the gauge group of the TSCC as in Eq.~\eqref{eq:G-tscc}.}
    \label{fig:CC}
\end{figure}

\subsection{Topological subsystem color codes}\label{subsec:TSCC}
Topological subsystem color codes (TSCCs) are a class of subsystem codes obtained from TCCs
\cite{bombin2010topological}.
To construct a TSCC, we construct a hypergraph from the 2-colex on which the color code is defined.
Recall that the hypergraph is defined by an ordered pair $(\mathsf{V}, \mathsf{E})$ where $\mathsf{V}$ is  the set of vertices of the hypergraph and $\mathsf{E}$, called edges of the hypergraph, is a collection of subsets of $\mathsf{V}$.
First, we replace every TCC vertex with a triangle. 
The three vertices of triangle form a rank-3 edge.
If $u,v,w$ are the vertices of the triangle, then the corresponding rank-3 edge is denoted 
$(u, v, w)$.
These newly formed vertices are denoted $\mathsf{V}$.
Next we split every  edge to two edges: solid and dashed, placed in alternating fashion around each triangle, see Fig.~\ref{fig:CC}. 
These newly formed edges are called rank-2 edges. 
The resulting object is a hypergraph as shown in \cite{suchara2011constructions}. 
The collection of rank-2 edges  is denoted $\mathsf{E}_2$ while the set of rank-3 edges
is denoted $\mathsf{E}_3$.
Together they constitute $\mathsf{E}=\mathsf{E}_2\cup \mathsf{E}_3$, the edges of the hypergraph.  
The hypergraph is now defined by the ordered pair $(\mathsf{V}, \mathsf{E})$.

Faces are somewhat more complicated to define. 
Each face of the color code gives rise to a face in the hypergraph. It also inherits the color 
from the underlying face in the 2-colex. 
We denote these faces by $\mathsf{F}$.
The  $c$ colored faces are denoted $\mathsf{F}_c$. 
Note that $\mathsf{F}=\mathsf{F}_r \cup \mathsf{F}_g\cup \mathsf{F}_b$. 

One can construct a subsystem code from the hypergraph obtained from the 2-colex
as follows, see \cite{bombin2010topological,suchara2011constructions}. 
Place qubits on the vertices of the hypergraph and for each rank-2 edge of the hypergraph 
we associate a Pauli operator. 
Every rank-3 edge $e=(u,v,w)$ is also assigned an operator $K_e=Z_u Z_v Z_w$.
Note that  for each  rank-3 edge $(u,v,w) $ we define three  $ZZ$ operators for each pair of vertices,
namely,  $Z_uZ_v$, $Z_vZ_w$ and $Z_uZ_w$.
Of these three $ZZ$ operators only two are independent.
These 2-body $ZZ$ operators along with the rank-2 edge operators generate the gauge group of the 
TSCC. 
(Note that the operator $Z_u Z_v Z_w$ is not a gauge operator.)
If a pair of vertices $e=(u,v)$ form a rank-2 edge or belong to a rank-3 edge we define
\begin{eqnarray}
K_e=K_{(u,v)} &= &\left\{\begin{array}{ll}
X_u X_v& (u,v) \text{ is dashed edge}\\
Y_u Y_v& (u,v) \text{ is solid edge}\\
Z_u Z_v& (u,v) \text{ is  in some hyperedge}
\end{array} \right.
\end{eqnarray}
We can then define the gauge group of the subsystem code as follows.
\begin{eqnarray}
\mathcal{G} & = & \langle K_{(u,v)} \mid (u,v) \text{ are adjacent} \rangle \label{eq:G-tscc}
\end{eqnarray}
The stabilizers and the logical operators of the TSCC are completely determined by the gauge group.
In case of TSCCs, they can be characterized in terms of  cycles of the hypergraph.

\begin{center}
\begin{figure}[H]
\tikzset{every picture/.style={line width=0.75pt}}
\centering
    \hspace*{-0.2cm}\includegraphics[scale=0.85]{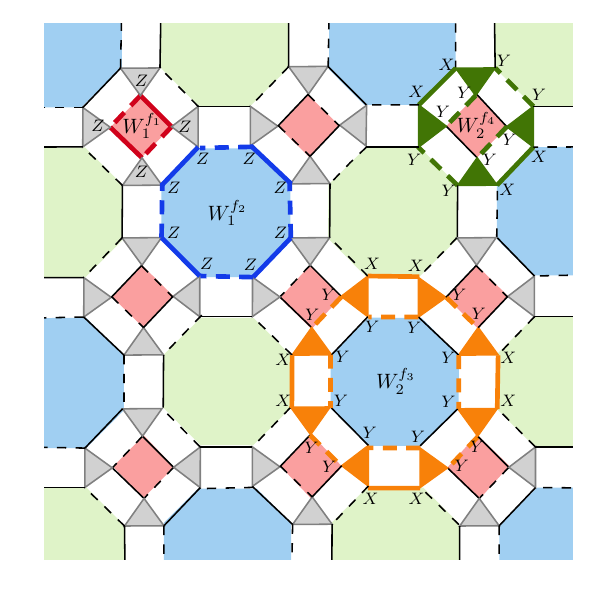} 
    \caption{Subsystem code defined on square octagon lattice. Also shown are the cycles and the associated stabilizers (in color). 
    Every face has a $Z$ type stabilizer coming from the rank-2 cycle, see $W_1^{f_1}$ and $W_1^{f_2}$. 
    Every face  also  has a stabilizer attached to a hypercycle, see  $W_2^{f_3}$ and $W_2^{f_4}$. 
    This stabilizer is neither $X$ type nor $Z$ type. 
    }
    \label{Fig:lattice+stab}
\end{figure}
\end{center}

A cycle in a hypergraph is a collection of edges such that every vertex has an even degree with respect to these edges. 
A rank-2 cycle involves only rank-2 edges. A hypercycle involves both rank-2 edges and hyperedges (rank-3 edges).
To every cycle $\sigma$ we can associate a operator as follows.
\begin{eqnarray}
W(\sigma) = \prod_{e\in \sigma}K_e \label{eq:hypercycle-op}
\end{eqnarray}
The importance of these operators is that they are precisely the operators in 
$C(\mathcal{G})$, the centralizer of $\mathcal{G}$.
A generating set for $C(\mathcal{G})$ can be given by considering all the cycles.
Cycles of trivial homology generate the stabilizer of the TSCC, as shown in Fig.~\ref{Fig:lattice+stab}.
Cycles of nontrivial homology give rise to the logical operators of the TSCC.

For every face $f \in \mathsf{F}$, we associate two independent cycles: 
  \begin{compactenum}[i)]
    \item A rank-2 cycle denoted $\sigma_1^f$. This is formed by the rank-2 edges in the boundary of $f$.
    \item A hypercycle denoted $\sigma_2^f$. This is formed by the rank-2 as well as rank-3 edges in the boundary of 
    $f$.
  \end{compactenum}
  (A dependent hypercycle can be generated by $\sigma_1^f$ and $\sigma_2^f$.) 
The hypercycle $\sigma_2^f$ is chosen so that the edges corresponding to the $XX$ type gauge operators in the boundary of $f$
are chosen to be in the hypercycle, see Fig.~\ref{Fig:lattice+stab}.
Cycles of trivial homology correspond to stabilizer generators of the subsystem code. 
We denote the stabilizer from rank-2 cycles as  $W_1^f$ and the stabilizer from the hypercycle as $W_2^f$.
\begin{subequations}
\begin{eqnarray}
W_1^f & = & \prod_{e\in \sigma_1^f} K_e  = \prod_{v \in f} Z_v  \label{eq:w1f-tscc}\\
W_2^f & = & \prod_{e\in \sigma_2^f} K_e = \prod_{(u ,v ,w )\in f} X_{u}Y_{v}Y_{w },\label{eq:w2f-tscc}
\end{eqnarray}
\end{subequations}
where $K_e$ is the edge operator associated to the edge $e$.
The alternate forms simply follow by substituting for the edge operators $K_e$.
Fig.~\ref{Fig:lattice+stab} shows these stabilizers: $W_1$ on faces $f_1$ and $f_2$ and  $W_2$ on faces $f_3$ and $f_4$.

The following dependencies exist among the stabilizer generators of the topological subsystem color codes on a torus. 
\begin{subequations}
\begin{eqnarray}
\prod_{f \in \mathsf{F}_r} W_2^f \prod_{f \in \mathsf{F}_g} W_2^f = \prod_{f\in \mathsf{F}_b} W_1^f \prod_{f\in \mathsf{F}_r} W_1^f \label{eq:tsc-stab-dep1}\\
\prod_{f \in \mathsf{F}_r} W_2^f \prod_{f \in \mathsf{F}_b} W_2^f = \prod_{f\in \mathsf{F}_g} W_1^f \prod_{f\in \mathsf{F}_b} W_1^f \label{eq:tsc-stab-dep2}
\end{eqnarray}
\end{subequations}

The above relations show that there are $s=2|\mathsf{F}|-2$ independent stabilizer generators for the 
TSCC assuming that there are $|\mathsf{F}|$ faces.

\begin{remark}\label{rem:tsc-params}
When defined on a torus we can show that the number of qubits is $n=6|\mathsf{F}|$.
The number of independent gauge generators is given by $2r+s=10|\mathsf{F}|-2$,
where $r$ is the number of gauge qubits which gives us 
$r=4|\mathsf{F}|$. 
The number of encoded qubits is $k=2$.
\end{remark}

\subsection{Mapping a  TSCC onto color codes}\label{sec:structure}
\begin{figure}[H]
    \centering
    \includegraphics[scale=0.75]{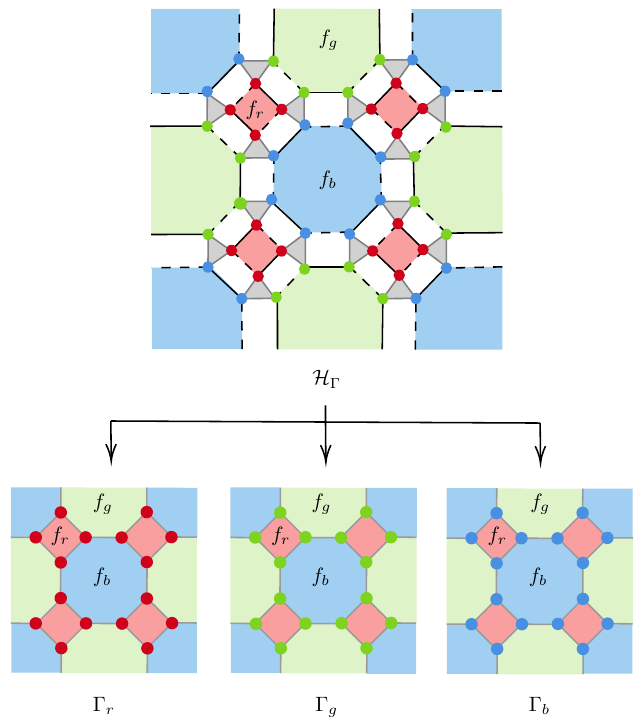}
    \caption{Faces of TSCC are colored in red, green and blue. Qubits are highlighted with the color of the face they belong to. Every stack of qubits of same color  form a color code identical to the parent color code of the TSCC.
    The faces in each stack are labeled same as the parent face in the hypergraph.}
    \label{fig:threecopies}
\end{figure}

Ref.~\cite{bombin2012universal}  proposed a mapping of subsystem codes onto three copies of color 
codes. 
We illustrate this in Fig.~\ref{fig:threecopies}, for the TSCC derived from the square octagon lattice.
The faces of the TSCC are colored with three different colors. 
Qubits belonging to faces of same color are grouped together and all of them are colored the same. 
Each of this stack (group) of qubits can be reinterpreted as a color code, see Fig.~\ref{fig:threecopies}.
This mapping could be viewed as a reinterpretation of the operators on the subsystem code in terms of  these copies of color codes. 

In mapping a TSCC to three copies of the parent TCC, we extend the stabilizer of the
subsystem code to $\mathcal{S}_e \subseteq \mathcal{G}$, where $\mathcal{S}_e$ is generated by all
the stabilizers of the three copies of the parent color code.
Therefore,
we have 
\begin{eqnarray}
\mathcal{S}\subseteq \mathcal{S}_e  \subseteq \mathcal{G} \subseteq C(\mathcal{G})\subseteq C(\mathcal{S}_e) \subseteq C(\mathcal{S}).\label{eq:tsc-tcc-stabilizer-reln}
\end{eqnarray}
Note that $\mathcal{S}_e$ is an Abelian group. 
There exist other ways to map a TSCC onto color codes, see \cite{bombin2012universal,gayatri2018decoding}.

\subsection{\label{subsec:errormodel}Error model}  
We assume the erasures occur independently on each qubit with a probability $\varepsilon$ and each qubit is erased with probability $\varepsilon$.
Once the loss is detected, the erased qubit is replaced with a completely mixed state. For the purposes of decoding, we can measure the syndrome with the replaced qubit \cite{moncy2018performance,kulkarni2019decoding,delfosse2020linear,delfosse2012upper}.
Since the completely mixed state can be written as 
     $\frac{I}{2}=\frac{1}{4}\rho + \frac{1}{4} X \rho X +\frac{1}{4} Y \rho Y + \frac{1}{4} Z \rho Z$
this is equivalent to placing one of the Pauli errors, $I, X, Y, Z$ with equal probability on the erased qubit. 
These errors are induced by the erasure and we refer to them as erasure induced Pauli errors. 
Decoding over an erasure channel requires us to correct the erasure induced Pauli errors. 
\section{Overview of the gauge fixing decoders}\label{sec:overview}
In this section we give an overview of proposed erasure decoders for TSCCs.
As discussed in the previous section, for correcting the erasure errors, we can map them to  Pauli errors on the erased qubits where each Pauli error occurs with equal probability. 
Although TSCCs are not CSS type codes, we can still correct these induced  $X$ and $Z$ errors separately as in the case of the decoders proposed for the depolarizing channel \cite{suchara2011constructions,gayatri2018decoding}. 
The proposed decoders have two stages: one for correcting  $X$ errors and one for correcting $Z$ errors.

For the proposed decoders we map TSCC to three color code copies (refer Section~\ref{sec:structure}) and decode over the color codes instead of decoding on the TSCC directly. 
Decoding by mapping onto the color codes implies that we augment the stabilizer by additional operators from the gauge group. 
We refer to this as gauge fixing since normally they are not constrained.
However, in the proposed decoders they are also measured and treated as stabilizers. 
(Note that the 2-body gauge generators are not treated as stabilizers.)
We propose two decoders which differ in how the phase flip errors are corrected. 

For the first decoder, bit flip errors are corrected by mapping the TSCC onto three copies of color codes. 
This requires additional gauge operators corresponding to the $Z$ type stabilizers of the color codes to be measured. 
These are used only for correcting the bit flip errors. 
After the bit flip errors are corrected the phase flip errors are corrected.
The $Z$ errors are  corrected using the stabilizers of the subsystem code (hypercycle stabilizers). 
The phase flip errors are corrected by mapping TSCC to the parent color code from which the subsystem color code was derived. 
We refer to this decoder as the {\em partial gauge fixing decoder}.

For the second decoder, the bit flip errors are corrected as in the first decoder. 
In addition, the phase flip errors are also corrected by mapping them to color codes
as in case of bit flip errors.
This uses the $X$ type stabilizers of the color codes also. 
These are used for correcting the phase flip errors. 
The stabilizers of the subsystem code corresponding to the hypercycles are not used directly in this decoder. 
The total number of gauge operators that are promoted to stabilizers is almost close to the maximal set possible. 
We refer to this as {\em order maximal gauge fixing decoder}.
However, this additional complexity leads to an improvement in performance as will be seen 
later in Section~\ref{sec:simulation_results}.

\begin{figure}[H]
    \centering
    \tikzset{every picture/.style={line width=0.75pt}} 

\begin{tikzpicture}[x=0.75pt,y=0.75pt,yscale=-1,xscale=1]

\draw (219.67,1688) node [anchor=north west][inner sep=0.75pt]  [color={rgb, 255:red, 0; green, 0; blue, 0 }  ,opacity=1 ] [align=left] {\begin{minipage}[lt]{114.47pt}\setlength\topsep{0pt}
\begin{center}
Syndrome measurement\\using Algorithm~\ref{alg:direct}
\end{center}
\end{minipage}};
\draw    (298.97,1753) -- (298.97,1725) ;
\draw [shift={(299,1753)}, rotate = 269.42] [fill={rgb, 255:red, 0; green, 0; blue, 0 }  ][line width=0.08]  [draw opacity=0] (8.93,-4.29) -- (0,0) -- (8.93,4.29) -- cycle    ;
\draw (203.67,1760) node [anchor=north west][inner sep=0.75pt]  [color={rgb, 255:red, 0; green, 0; blue, 0 }  ,opacity=1 ] [align=left] {\begin{minipage}[lt]{136.61pt}\setlength\topsep{0pt}
\begin{center}
Preprocessing using peeling \\and clustering
\end{center}
\end{minipage}};

\draw    (298.97,1822) -- (298.97,1794) ;
\draw [shift={(299,1822)}, rotate = 269.42] [fill={rgb, 255:red, 0; green, 0; blue, 0 }  ][line width=0.08]  [draw opacity=0] (8.93,-4.29) -- (0,0) -- (8.93,4.29) -- cycle    ;

\draw (246.33,1832) node [anchor=north west][inner sep=0.75pt]   [align=left] {\begin{minipage}[lt]{77.59pt}\setlength\topsep{0pt}
\begin{center}
Stage 1: X error \\correction using\\Algorithm~\ref{alg:dec_using_tcc_synd}
\end{center}
\end{minipage}};

\draw (237.83,1890.67) node [anchor=north west][inner sep=0.75pt]  [color={rgb, 255:red, 0; green, 0; blue, 0 }  ,opacity=1 ] [align=left] {\begin{minipage}[lt]{88.97pt}\setlength\topsep{0pt}
\begin{center}
\textcolor[rgb]{0.19,0.28,0.76}{Decoding on three }\\\textcolor[rgb]{0.19,0.28,0.76}{copies of }\\\textcolor[rgb]{0.19,0.28,0.76}{color codes}
\end{center}
\end{minipage}};
\draw    (298.97,1943.33) -- (298.97,1972) ;
\draw    (410.5,1972) -- (298.97,1972) ;
\draw    (298.97,1972) -- (196.5,1972) ;

\draw    (196.5,1972) -- (196.5,2003) ;
\draw [shift={(196.5,2003)}, rotate = 270.08] [fill={rgb, 255:red, 0; green, 0; blue, 0 }  ][line width=0.08]  [draw opacity=0] (8.93,-4.29) -- (0,0) -- (8.93,4.29) -- cycle    ;
\draw    (410.5,1972) -- (410.5,2003) ;
\draw [shift={(410.46,2003)}, rotate = 270.08] [fill={rgb, 255:red, 0; green, 0; blue, 0 }  ][line width=0.08]  [draw opacity=0] (8.93,-4.29) -- (0,0) -- (8.93,4.29) -- cycle    ;

\draw (155,1934) node [anchor=north west][inner sep=0.75pt]  [font=\small,color={rgb, 255:red, 0; green, 0; blue, 0 }  ,opacity=1 ] [align=left] {\begin{minipage}[lt]{66.27pt}\setlength\topsep{0pt}
\begin{center}
\textcolor[rgb]{0,0,0}{Partial gauge }\\\textcolor[rgb]{0,0,0}{fixing decoder}
\end{center}
\end{minipage}};

\draw (335,1934) node [anchor=north west][inner sep=0.75pt]  [font=\small,color={rgb, 255:red, 0; green, 0; blue, 0 }  ,opacity=1 ] [align=left] {\begin{minipage}[lt]{103.69pt}\setlength\topsep{0pt}
\begin{center}
\textcolor[rgb]{0,0,0}{Order maximal gauge }\\\textcolor[rgb]{0,0,0}{fixing decoder${}^\dagger$}
\end{center}
\end{minipage}};

\draw (145,2015) node [anchor=north west][inner sep=0.75pt]   [align=left] {\begin{minipage}[lt]{77.02pt}\setlength\topsep{0pt}
\begin{center}
Stage 2: Z error \\correction using\\Algorithm~\ref{alg:partial_z_correction}
\end{center}
\end{minipage}};
\draw (135,2075) node [anchor=north west][inner sep=0.75pt]  [color={rgb, 255:red, 0; green, 0; blue, 0 }  ,opacity=1 ] [align=left] {\begin{minipage}[lt]{91.81pt}\setlength\topsep{0pt}
\begin{center}
\textcolor[rgb]{0.19,0.28,0.76}{Decoding on parent}\\\textcolor[rgb]{0.19,0.28,0.76}{color code}
\end{center}
\end{minipage}};

\draw (356.5,2015) node [anchor=north west][inner sep=0.75pt]   [align=left] {\begin{minipage}[lt]{77.02pt}\setlength\topsep{0pt}
\begin{center}
Stage 2: Z error \\correction using\\Algorithm~\ref{alg:dec_using_tcc_synd}
\end{center}
\end{minipage}};

\draw (342,2075) node [anchor=north west][inner sep=0.75pt]  [color={rgb, 255:red, 0; green, 0; blue, 0 }  ,opacity=1 ] [align=left] {\begin{minipage}[lt]{98.59pt}\setlength\topsep{0pt}
\begin{center}
\textcolor[rgb]{0.19,0.28,0.76}{Decoding on three }\\\textcolor[rgb]{0.19,0.28,0.76}{copies of color codes}
\end{center}

\end{minipage}};

\end{tikzpicture}
    \caption{Overview of proposed two stage subsystem decoders over the erasure channel. ${}^\dagger$For the second decoder, the order of Stage 1 and Stage~2 can be swapped or implemented parallely.}
    \label{fig:overview2}
\end{figure}
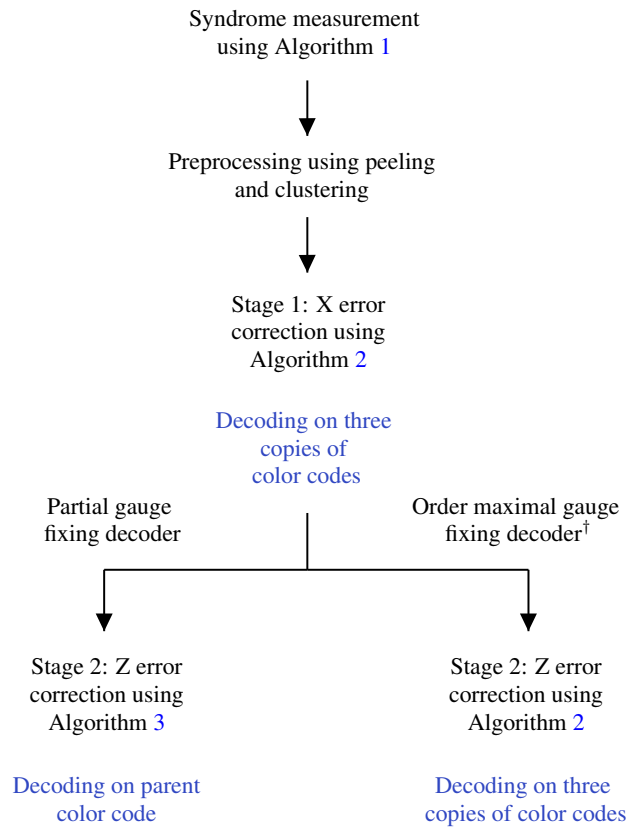

\subsection{Gauge fixing} \label{subsec:tscc-2-tccs}
In mapping the subsystem code to the color codes, the stabilizer of the subsystem code is extended to 
a larger stabilizer. 
This leads to gauge fixing where some of the gauge operators are promoted to checks \cite{vuillot2019code}. 
Since our original code is a subsystem code, and we plan to decode it using the three stacks of color codes,
a careful counting of the stabilizers on the color codes leads to the following observations. 
First of all the code we are interested in decoding is the subsystem code.
This subsystem  code  has $2|\mathsf{F}|$ stabilizer generators, with two dependencies. 
On the other hand mapping the subsystem color code to color code requires 
us to use $2|\mathsf{F}|$ generators for each color code i.e. $6|\mathsf{F}|$ stabilizer generators. 
(For correcting the $X$ type errors we only need $3|\mathsf{F}|$ syndromes and for correcting the phase flip errors
$3|\mathsf{F}|$).
This means that we must be able generate additional ``syndromes'' on the color codes 
consistently. 
This naturally leads us to the idea of gauge fixing where in addition to the stabilizer generators of the subsystem code, we also measure some additional commuting set of gauge operators.

To obtain the additional syndromes on the color codes, we measure them directly on TSCC.
We further show that there exists a decomposition of the color code stabilizers in smaller body gauge generator measurements, but measurement of these individual gauge generators may not be possible due to their anticommuting nature. 

We summarize the key requirements on the measurement of syndromes 
for the proposed decoders in light of the above  discussion on gauge fixing.

\begin{compactenum}[M1.]
    \item Completeness: In order that the syndromes of the color codes are generated from the gauge measurements, we 
also require that all their stabilizers be generated gauge operators.
\item Locality: From the point of view of fault tolerance we want the stabilizers of color codes should also be measured locally on the TSCC.
\item Sequencing: Any gauge operator to be measured must commute with the product of previously measured gauge operators.
\end{compactenum}

\section{Syndrome measurement and preprocessing}\label{sec:synd_and_prepro}
A popular approach for decoding topological codes is to map the original code into some other quantum code for which efficient decoders are known \cite{bombin2012universal,nautrup2017fault,aloshious2019erasure,haah2016algebraic,delfosse2014decoding}.
For our decoding algorithm, we use the map discussed in Sec.~\ref{sec:structure}.
Instead of correcting errors on the TSCC, we correct the copies of color codes and then lift the error to subsystem code.
In Sec.~\ref{subsec:measurement_of_TCC} we discuss the measurement for TCC and TSCC stabilizers.
In Sec.~\ref{subsec:preprocess} we discuss the preprocessing techniques to do before the actual decoding steps.

\subsection{Measurement of TCC and TSCC stabilizers}\label{subsec:measurement_of_TCC}
We map the TSCC onto three (identical) copies of color codes.
Our plan is to decode on the color codes instead on decoding on the TSCC.
For this we need to measure the color code stabilizers on the TSCC.

\begin{center}
\begin{figure}[H]
    \centering
    \includegraphics[scale=0.67]{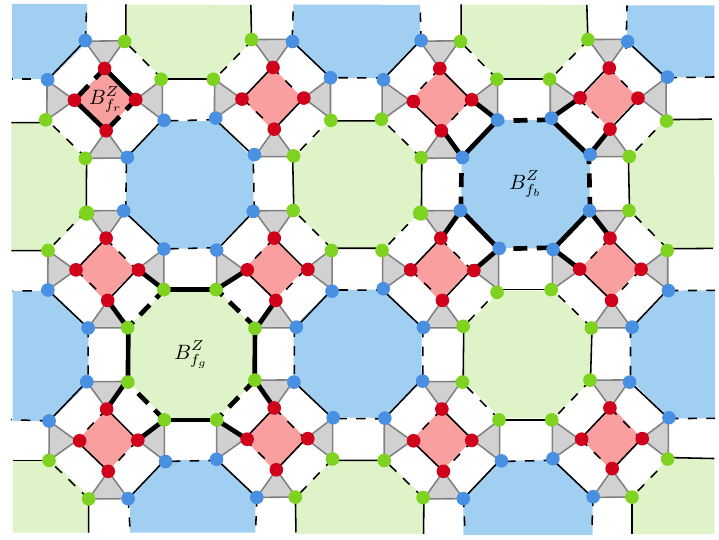}
    \caption{Bold edges on the TSCC denote the 2-body decomposition for $B_{f_r}^Z$, $B_{f_b}^Z$ and $B_{f_g}^Z$  stabilizers for color code on red stack.}
    \label{fig:All_redZstab}
\end{figure}
\end{center}

\begin{center}
\begin{figure}[H]
    \centering
    \includegraphics[scale=0.67]{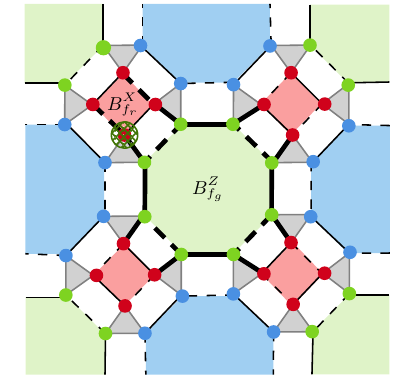}
    \caption{Bold edges on the TSCC denote the 2-body decomposition for $B_{f_r}^X$ and $B_{f_g}^Z$  stabilizers for color code on red stack. The highlighted green net patterned circle shows the qubit where the 2-body $XX$ gauge generator and the 2-body $ZZ$ gauge generator anticommute.  
    }
    \label{fig:anti_comm}
\end{figure}
\end{center}

One idea to reduce the complexity of the measurements of color coder stabilizers is to break them down to smaller body measurements (product of gauge generators).
This is similar to how a TSCC stabilizer is measured using measurements of commuting gauge generators and then combining their outcome.
Fig.~\ref{fig:All_redZstab} illustrates how the $Z$ type stabilizers on the red stack can be decomposed as product of gauge generators. 
We can obtain similar map for $X$ type stabilizers.
Measurement of all these operators  using only 2 body gauge generators is not possible since they anticommute among each other. 
Fig.~\ref{fig:anti_comm} shows an example of how two gauge generators of stabilizers $B_{f_r}^X$ and $B_{f_g}^Z$, on the red stack, anticommute with each other at the highlighted location. Due to this, both the stabilizers cannot be measured simultaneously using the 2-body gauge operators\footnote{We would like to acknowledge Oscar Higgot for highlighting this issue in the previous version of this paper.}.

\begin{center}
\begin{figure}[H]
    \centering
    \includegraphics[scale=0.67]{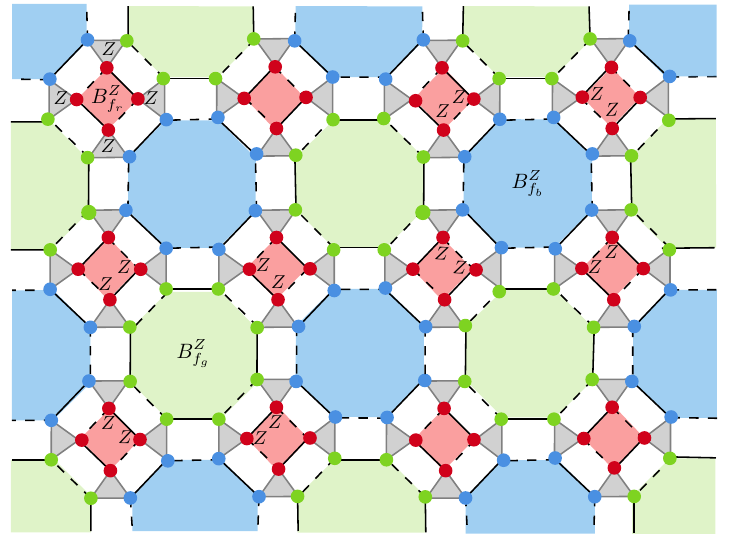}
    \caption{Direct measurement of 4-body $ B_{f_r}^Z$ stabilizer and 8-body $B_{f_b}^Z$ and $B_{f_g}^Z$ stabilizers for color code on red stack.}
    \label{fig:All_redZstab_direct}
\end{figure}
\end{center}

Hence instead of measuring the color code stabilizer using product of individual gauge generators, we measure 4-body or 8-body TCC stabilizer directly on TSCC.
Fig.~\ref{fig:All_redZstab_direct} shows how these measurements are done on the TSCC to obtain the $Z$ type color code stabilizers on red stack.

Similarly, we can directly measure both $X$ type and $Z$ type stabilizers, for all three stacks, on the TSCC using 4 and 8-body measurements.
Note that while measuring stabilizers for $c$ stack, where $c \in \{r,g,b\}$, only the qubits on $c$ face of TSCC are involved. 
\begin{remark}
Note that all the stabilizer generators of the color codes on each of the three stacks are generated as elements of the gauge group of the TSCC.
The dependencies among the individual color codes are respected by this decomposition. 
So for instance on the color code on each stack the dependencies corresponding to Eq.~\eqref{eq:tcc-z-stab-dep} 
are also respected. 
\end{remark}
The above remark makes sure that we get the correct measurement outcome for all the stabilizers after the syndrome measurement process.
The syndrome measurement is given in Algorithm~\ref{alg:direct}.

Observe that every $W_1^f$ stabilizer corresponds to a color code stabilizer.
In fact, $(B_{f}^Z)_c=W_1^{f}$ where $f\in \mathsf{F}_c$.
On the other hand, none of the color code stabilizer directly corresponds to the hypercycle stabilizer $W_2^f$ of TSCC since $W_2^f$ have support on all three stacks. 
\begin{figure}
    \centering
    \includegraphics[scale=0.75]{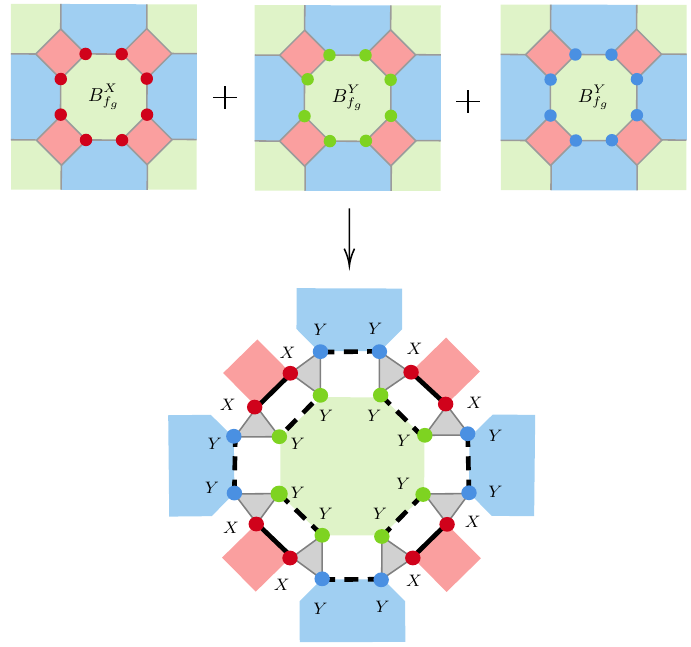}
    \caption{Decomposition of $W_2^{f_g}$ into the stabilizers on color codes. 
    namely,  $(B_{f_{g}}^{X})_r$, $(B_{f_{g}}^{Y})_g$, and $(B_{f_{g}}^{Y})_b$.
    }
    \label{fig:sum_to_hypercycle}
\end{figure}
However, the hypercycle stabilizer $W_2^f$ can be decomposed into three stabilizers each supported on a stack of different color, see  Fig.~\ref{fig:sum_to_hypercycle}.
The figure shows how $X$ type stabilizer on the red stack  and $Y$ type stabilizers on the blue and green stacks  can be combined to form the hypercycle stabilizer on green face of TSCC. 
\begin{subequations}
\begin{eqnarray}
  W_2^{f_r} &= & {(B_{f_{r}}^{Y})}_r  {(B_{f_{r}}^{Y})}_g  {(B_{f_{r}}^{X})}_b  \\
  W_2^{f_g} &=&{(B_{f_{g}}^{X})}_r {(B_{f_{g}}^{Y})}_g  {(B_{f_{g}}^{Y})}_b  \\
  W_2^{f_b} &= & {(B_{f_{b}}^{Y})}_r {(B_{f_{b}}^{X})}_g  {(B_{f_{b}}^{Y})}_b 
\end{eqnarray} \label{eqn:hypercycle-decompo}
\end{subequations}
Using these decompositions, we can also measure the syndromes of the TSCCs using the color code stabilizer outcomes.
\begin{algorithm}[H]
\caption{Syndrome computation for TCC and TSCC}
\begin{algorithmic}[1]
    \REQUIRE {TSCC hypergraph $\mathcal{H}_{\Gamma}$, set of erasure positions $\mathcal{E}$.}
    \ENSURE {$X$ error syndrome $s_c^X$ and $Z$ error syndrome $s_c^Z$ for the color code on the $c$-stack and syndromes of TSCC.}
    \STATE Group the qubits according to color of faces on TSCC.
    \FOR {every $c$-stack, $c \in \{r,g,b\}$ }
    \STATE Measure 4-body $(B_{f_r}^Z)_c$ and $(B_{f_r}^X)_c$ stabilizers on TSCC.
    \STATE Measure 8-body $(B_{f_g}^Z)_c$, $(B_{f_g}^X)_c$, $(B_{f_b}^Z)_c$ and $(B_{f_b}^X)_c$ stabilizers on TSCC.
    \ENDFOR
    \STATE Calculate $W_2^{f}$ using Eq.~\eqref{eqn:hypercycle-decompo} for all $f\in \mathsf{F}$.
\STATE Return $s_c^Z$ and $s_c^X$, $c \in \{r,g,b\}$, syndromes of $W_1^f$ and $W_2^f$. \COMMENT{Note that $(B_{f}^Z)_c=W_1^{f}$ where $f\in \mathsf{F}_c$. }
\end{algorithmic}
\label{alg:direct}
\end{algorithm}

\subsection{Preprocessing the TSCC}\label{subsec:preprocess}
Algorithm~\ref{alg:direct} shows how all the color code stabilizer and TSCC stabilizer measurements are obtained.
In this section we state how the TSCC syndrome can be used to do preprocessing on the TSCC.
We use two preprocessing techniques, clustering and peeling for our algorithms.
TSCC syndromes are used during the peeling stage.
First, we apply the preprocessing techniques to remove some simple erasure patterns.
We update the TCC syndromes when any correction is done during peeling.
After peeling we perform clustering of the remaining erasures.
For more details, refer \cite{ITW_TSCC}.

\section{First stage: $X$ error correction using gauge fixing}\label{sec:first_stage}
In this section we study the first stage- bit flip error correction.
As shown in Fig.~\ref{fig:threecopies}, we map the TSCC to three copies of color codes.
Since there is one to one correspondence between the qubits of the color codes and the TSCC, 
we can directly map the erasures on TSCC onto the color codes.
Next we obtain the $X$ error syndrome, by measuring $Z$ type stabilizers of the color codes as shown in Fig.~\ref{fig:All_redZstab_direct}.
Once the syndrome on each stack is obtained, we decode the color code on that stack using a color code erasure decoder and get an error estimate.
We adapt the color code erasure decoder proposed in \cite{aloshious2019erasure}.
The final step is to lift the estimate from all the color codes to the TSCC.
Since every qubit of the color code is a dedicated TSCC qubit (see Sec.~\ref{sec:structure}), we lift the error estimate at the same location on the TSCC.
Note that since no stack share any qubit, all the three color codes can be decoded parallely and independently.

We also update the hypercycle syndrome $W_2^f$ at the end of this stage.
This is done because correcting the $X$ errors clears $B_f^Z$ stabilizers and hence modifies the $B_f^Y$ stabilizers.
As shown in Eqn.~\ref{eqn:hypercycle-decompo}, the hypercycle stabilizer directly depends on both $X$ and $Y$ type color code stabilizers.
Hence any modification in the color code stabilizer syndrome also affects the hypercycle syndrome.

The complete decoding procedure for the first stage is given in Algorithm~\ref{alg:dec_using_tcc_synd} with $\sigma= X$.
During this stage of correction, we clear all the $Z$ type stabilizers of the color codes. 
Note that only a subset of these stabilizers are actually a stabilizer on the TSCC. (Recall that $(B_{f}^Z)_c=W_1^{f}$ where $f\in \mathsf{F}_c$.) 
Hence we are fixing additional TSCC gauge operators, on top of TSCC rank-2 stabilizers, while clearing the bit flip errors.

\begin{algorithm}[H]
\caption{TSCC erasure decoder via mapping to TCCs: Bit flip (Phase flip) error correction}
\begin{algorithmic}[1]
    \REQUIRE {TSCC hypergraph $\mathcal{H}_{\Gamma}$, set of erasure positions $\mathcal{E}$ and $\sigma$ error syndrome $s_c^\sigma$, $c \in \{r,g,b\}$. \COMMENT{Note $\sigma$ can be either $X$ or $Z$}}
    \ENSURE {$\sigma$ error estimate $\widehat{E}^\sigma$}
    \STATE $\widehat{E}^\sigma = I$
\STATE Group the qubits according to color of faces on TSCC.
    \FOR {every $c$-stack, $c \in \{r,g,b\}$ }
        \STATE $\widehat{E}_c = I$
        \STATE Map erasure locations directly as per the color.
        \STATE Given $s_c^\sigma$, decode color code using any color code erasure decoder to get estimate $\widehat{E}_c$.         
        \STATE Lift the estimate $\widehat{E}_c$ to the subsystem code. 
        \STATE Update $W_2^f$ syndrome according to the estimate. 
        \STATE Update $\widehat{E}^\sigma = \widehat{E}^\sigma \widehat{E}_c$.
    \ENDFOR
    \STATE Return $\widehat{E}_\sigma$ as the final  $\sigma$ error estimate.
\end{algorithmic}
\label{alg:dec_using_tcc_synd}
\end{algorithm}

\subsection{Residual errors after correcting bit flip errors }\label{subsec:residual-errors}
\begin{figure*}[ht] 
\begin{subfigure}[h]{0.4\textwidth}
    \centering
    \includegraphics[scale=0.5]{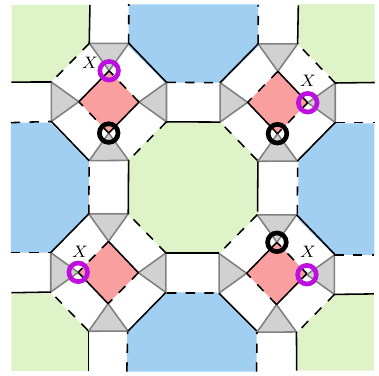}
    \subcaption{Original error on TSCC}
    \label{subfig:ITW_a}
\end{subfigure}
\begin{subfigure}[h]{0.56\textwidth}
    \centering
    \includegraphics[scale=0.5]{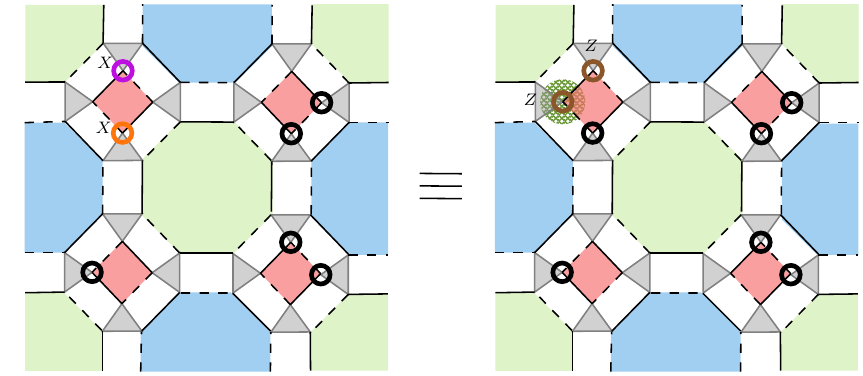}
    \subcaption{After applying $X$ correction operator based on $W_1^f$ syndrome.}
    \label{subfig:ITW_b}
\end{subfigure}
\captionsetup{justification=justified}
\caption{Fig.~\ref{subfig:ITW_a} shows the original error on TSCC. The purple rings denote erased locations with $X$ error and black rings denote erased locations with identity. On performing bit flip  correction as per \cite{ITW_TSCC}, the resultant TSCC is shown in Fig.~\ref{subfig:ITW_b}. The orange ring denotes the $X$ error estimate and brown ring denotes $Z$ error. As highlighted with green net patterned circle, there exists a residual $Z$ error on unerased qubit as well.}
\label{fig:ITW_example}
\end{figure*}

As can be seen from Algorithm~\ref{alg:dec_using_tcc_synd}, we decode the erasure 
induced bit flip errors by means of a color code erasure decoder. 
If the decoder only estimates on the erased qubits, then one can immediately proceed to correct the  phase error caused by the erasure.
However, in some cases, this decoder can cause a new $Z$ type error on unerased qubits as in the case of \cite{ITW_TSCC}, see for example Fig.~\ref{fig:ITW_example}.
So this must be accounted for in the next stage. 
In this paper we used the erasure color code decoder proposed in \cite{aloshious2019erasure}.
This particular decoder can potentially return an error estimate on nonerased qubits up to an $X$
type stabilizer or logical operator on the color code. 
(This is because the decoder works by mapping on to surface codes which can generate additional errors, see \cite{aloshious2019erasure} for more details.)
When the estimate is up to a stabilizer on the color code, on the subsystem code the error is up to an $X$ type gauge operator because $X$ type stabilizers on the color code are gauge operators on the subsystem code. 
When the estimate is up to a logical operator on the color code, on the subsystem code the error can either be a logical operator or an $X$ type gauge operator.

\begin{figure*}[ht] 
\captionsetup[subfigure]{justification=centering}
\begin{subfigure}[b]{0.19\textwidth}
    \centering
    \includegraphics[scale=0.5]{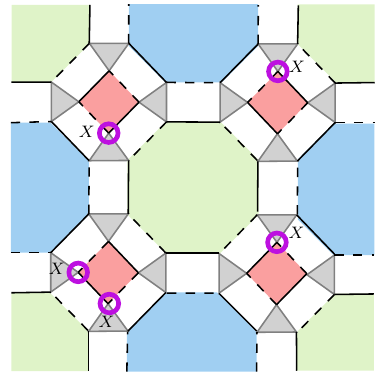}
    \subcaption{Original error on TSCC}
    \label{subfig:aa}
\end{subfigure}
\begin{subfigure}[b]{0.19\textwidth}
    \centering
    \includegraphics[scale=0.6]{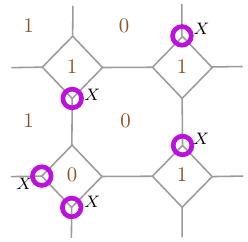}
    \subcaption{Error and syndrome on red color code}
    \label{subfig:bb}
\end{subfigure}
\begin{subfigure}[b]{0.19\textwidth}
    \centering
 \includegraphics[scale=0.6]{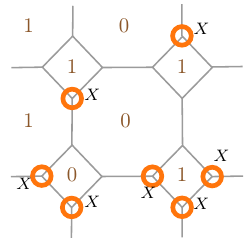}
    \subcaption{Error estimate on red color code}
    \label{subfig:cc}
\end{subfigure}
\begin{subfigure}[b]{0.19\textwidth}
    \centering
     \includegraphics[scale=0.6]{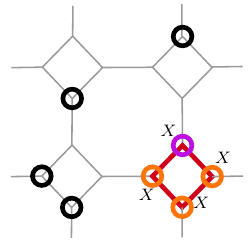}
    \subcaption{Effective error on red color code; a stabilizer}
    \label{subfig:dd}
\end{subfigure}
\begin{subfigure}[b]{0.19\textwidth}
    \centering
     \includegraphics[scale=0.5]{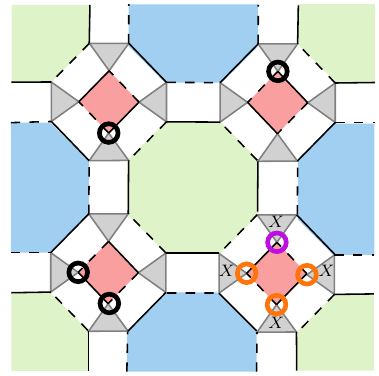}
    \subcaption{Lifting error estimate on TSCC}
    \label{subfig:ee}
\end{subfigure}
\caption{Original error on TSCC is shown in Fig.~\ref{subfig:aa}. Purple rings denote the erased locations with $X$ error. Fig.~\ref{subfig:bb} shows the error and syndrome on the red color code. In Fig.~\ref{subfig:cc}, orange ring denote the $X$ error estimate on red color code. Fig.~\ref{subfig:dd} shows the error $+$ error estimate on red color code. As highlighted with red, the resultant is a color code stabilizer. Fig.~\ref{subfig:ee} shows how after applying the color code estimate we get resultant $X$ type gauge operator.
}
\label{fig:example_partial_stabilizer}
\end{figure*}

Suppose a part of the error pattern is as shown in Fig.~\ref{subfig:aa}.
Note that support of the error lies only on the red color code, see Fig.~\ref{subfig:bb}. 
We measure the color code stabilizers directly on the TSCC to obtain the syndrome shown in  Fig.~\ref{subfig:bb}. 
With the syndrome and erased locations, we decode the color code using color code erasure decoder and obtain error estimate as shown in Fig.~\ref{subfig:cc}.
Note that the estimate is not same as the original  error, but it results in the same syndrome.
When we apply this estimate to the original error on the color code, it results in  a stabilizer as shown in Fig.~\ref{subfig:dd}. 
(Therefore, the correction is up to a stabilizer.)
When the estimate is lifted to the TSCC, we get a resultant $X$ type gauge operator, shown in Fig.~\ref{subfig:ee}.
This example shows that the final outcome can induce an $X$ type gauge error but not any additional $Z$ errors
on the unerased qubits.
The following lemma proves this result formally. 

\begin{figure*}[ht] 
\captionsetup[subfigure]{justification=centering}
\begin{subfigure}[b]{0.19\textwidth}
    \centering
     \includegraphics[scale=0.5]{Figures/ITW_a.pdf}
    \subcaption{Original error on TSCC}
    \label{subfig:a}
\end{subfigure}
\begin{subfigure}[b]{0.19\textwidth}
    \centering
     \includegraphics[scale=0.6]{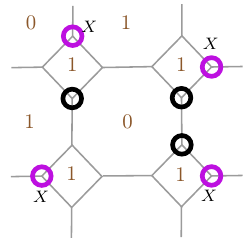}
    \subcaption{Error and syndrome on red color code}
    \label{subfig:b}
\end{subfigure}
\begin{subfigure}[b]{0.19\textwidth}
    \centering
 \includegraphics[scale=0.6]{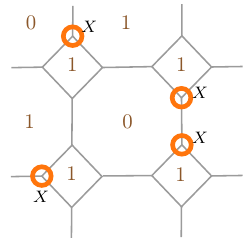}
    \subcaption{Error estimate on red color code}
    \label{subfig:c}
\end{subfigure}
\begin{subfigure}[b]{0.19\textwidth}
    \centering
     \includegraphics[scale=0.6]{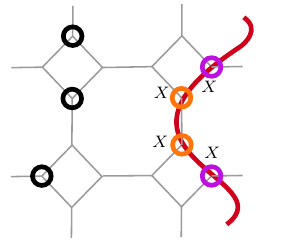}
    \subcaption{Effective error is a logical operator}
    \label{subfig:d}
\end{subfigure}
\begin{subfigure}[b]{0.19\textwidth}
    \centering
     \includegraphics[scale=0.5]{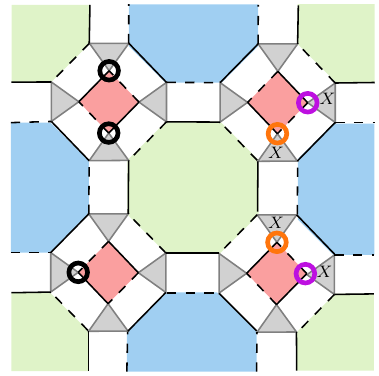}
    \subcaption{After applying error estimate on TSCC}
    \label{subfig:e}
\end{subfigure}
\caption{The original error on TSCC is shown in Fig.~\ref{subfig:a} (same as Fig.~\ref{subfig:ITW_a}). Fig.~\ref{subfig:b} shows the error and syndrome on color code on red stack. In Fig.~\ref{subfig:c}, orange ring denote the $X$ error estimate. Fig.~\ref{subfig:d} shows the error $+$ error estimate on the color code. Fig.~\ref{subfig:e} shows how after applying the color code estimate we get resultant $X$ type gauge operator.
}
\label{fig:example_partial_logical}
\end{figure*}

\begin{lemma}\label{lemma:no_Z_error}
On performing $X$ error correction on a TSCC by decoding on three copies of color codes through Algorithm~\ref{alg:dec_using_tcc_synd}, 
the lifted error estimate can either be an $X$ type gauge operator or a logical operator (bare or dressed).
If the estimate on the color codes is up to a stabilizer, then the lifted error estimate
does not contain any $Z$ error on the unerased qubits. 
\end{lemma}

\begin{proof}
Suppose that the $X$ part of the error on the TSCC is denoted $E$ and its restriction to the color code on $c$-stack
be $E_c$.
Let the error estimate returned by Algorithm~\ref{alg:dec_using_tcc_synd} be  $\widehat{E}_c$
on the color code on the $c$-stack. 
This estimate can be decomposed as $\widehat{E}_c=E_cLS$, where $L$ is a logical operator and $S$ is a stabilizer
of the color code. 
This estimate will be a pure $X$ type operator on the color code as it is a CSS code. 
When $\widehat{E}_c$ is equal to $E$, we get a perfect correction step and there is no residual error on the color code
or the TSCC. 
If $\widehat{E}_c$ is not equal to $E$, the estimate is up to a $X$ stabilizer or a logical operator. 
When this estimate is lifted to the TSCC, it can either be a logical operator or a pure $X$ type gauge operator. 
If the estimate is up to a (nontrivial) logical operator on the color codes, then from Eq.~\eqref{eq:stab-gauge-reln}, we see
that the lifted error estimate 
on the subsystem code, is an $X$ type gauge operator or a dressed logical operator of the subsystem code.
Recall that a dressed logical operator is a logical operator of the TSCC augmented with additional gauge generators.

In case the estimate is up to a stabilizer on the color codes, then on the subsystem code, the estimate is up to an 
$X$ type gauge operator that does not contain any $Z$ errors on the unerased qubits. 
This is because the ($X$ type) stabilizers of the color codes on each stack are in the gauge group. 
\end{proof}
Lemma~\ref{lemma:no_Z_error} shows that the lifted estimate on the TSCC does not result in any nonzero syndrome on the TSCC, 
therefore the second stage can also focus only on the erased qubits and ignore the unerased qubits.

Note that when the decoders on the color codes fail, it does not always lead to 
logical error on the TSCC as well.
The following example illustrates this. 
Consider now the same error pattern as Fig.~\ref{subfig:ITW_a} this time decoded using Algorithm~\ref{alg:dec_using_tcc_synd}.
For simplicity we choose an error pattern which is supported only on the red faces,
Fig.~\ref{subfig:a} shows the error. 
Fig.~\ref{subfig:b} shows the error pattern and the corresponding syndrome on the red stack. 
Suppose the decoder results in the estimate shown in Fig.~\ref{subfig:c}. 
On correcting according to this  estimate, the resultant operator is a logical operator on the color code as shown in Fig.~\ref{subfig:d}.
When the estimate shown in Fig.~\ref{subfig:c} is lifted to the TSCC, we get an $X$ type gauge operator as shown in Fig.~\ref{subfig:e}.
Estimate on color code up to a logical error can result to an $X$ type gauge operator on the TSCC or a dressed logical operator on TSCC. 
In either case it does not create any additional $Z$ error syndrome on the TSCC.

Before we conclude this section, in Table~\ref{tab:diffw-itw} we summarize the differences between the bit flip error 
decoding in  Ref.~\cite{ITW_TSCC} and Algorithm~\ref{alg:dec_using_tcc_synd}.

\begin{table}[H]
\begin{tabular}{l|l|p{2.9cm}}
\hline
\multicolumn{1}{c|}{\textbf{Parameters} }& \multicolumn{1}{c|}{\textbf{Ref.~\cite{ITW_TSCC} }} & \multicolumn{1}{c}{\textbf{Algorithm~\ref{alg:dec_using_tcc_synd}}}  \\ \hline
Type of decoding& Local  & Global   \\ \hline
\begin{tabular}[c]{@{}l@{}}$X$ error \\ correction on\end{tabular}    & TSCC   & Three copies of color codes \\ \hline
\begin{tabular}[c]{@{}l@{}}Error estimate \\ on each face\end{tabular} & Single $X$ error   & One or more $X$ errors  \\ \hline
Error estimate  & \begin{tabular}[c]{@{}l@{}}Can create residual \\  $Z$ error on \\ unerased qubits \end{tabular} & \begin{tabular}[c]{@{}l@{}} Estimate is a $X$ type  \\  gauge operator or  \\  logical operator\end{tabular} \\ \hline
\end{tabular}
\caption{Differences between first stage (bit flip error correction) of Ref.~\cite{ITW_TSCC} decoder and proposed decoder}\label{tab:diffw-itw}
\end{table}

\section{Second stage: $Z$ error correction}\label{sec:second_stage}

After performing $X$ error correction, the next step is to correct the $Z$ errors.
We corrected the $X$ errors with gauge fixing since we also fixed additional TCC stabilizers apart from TSCC stabilizers.
In this section, we propose two  algorithms  to correct the $Z$ errors.
One of them uses only the stabilizers of the TSCC while the other uses the TCC $X$ type stabilizers. 

\subsection{Correcting $Z$ errors without gauge fixing}\label{subsec:partial}
Recall that the TSCC is constructed from a $2$-colex using vertex expansion. 
For $Z$ error correction, we take the hypercycle syndromes of the TSCC and map them onto the parent $2$-colex $\Gamma$ from which the TSCC is derived. 
For face $f$, the hypercycle syndrome $W_2^f$ corresponds to the $X$ type stabilizer syndrome of the corresponding face on the $2$-colex $\Gamma$. 
Since we have already cleared the $X$ errors in the first stage,  the residual syndrome on
the lattice due to hypercycle stabilizers can be explained purely in terms of 
$Z$ errors alone. 
Next we also map the erasures from the TSCC to the parent 2-colex. 
A vertex of the $2$-colex is erased if any one or more qubits of the hyperedge  (inflated triangle) in the TSCC are erased. 
The syndromes of the hypercycles are projected  to the $2$-colex.
(Note that the hypercycle syndromes were updated after performing $X$ error correction.)
Once the  erased  locations  and syndromes are obtained on the underlying color code, we can adapt any color code erasure decoder to decode it.

After decoding, we  lift  the  error  estimate  on  the TSCC.
A $Z$ error estimate on a color code qubit is lifted to $Z$ error on any one of the qubits on the corresponding hyperedge in the TSCC.
The complete procedure for $Z$ error correction is given in Algorithm~\ref{alg:partial_z_correction}.

\begin{algorithm}[H]
\caption{Correcting $Z$ errors without gauge fixing}
\begin{algorithmic}[1]
    \REQUIRE {TSCC hypergraph $\mathcal{H}_{\Gamma}$ , erasure positions $\mathcal{E}$ and syndrome of hypercycle stabilizers ($W_2^f$).}
    \ENSURE {$Z$ Error estimate $\widehat{E}^Z$}
    \STATE Project the updated $W_2^f$ syndromes (from Algorithm~\ref{alg:dec_using_tcc_synd}) to the underlying parent color code of the TSCC.
    \STATE Map the erasure locations.
    \STATE Adapt any color code erasure decoder to obtain an error estimate $\widetilde{E}$ (on the color code). 
    \STATE Lift the error estimate $\widetilde{E}$ to the TSCC. Denote the lifted estimate $\widehat{E}^Z$.
    \STATE Return $\widehat{E}^Z$ as the error estimate for $Z$ errors.
\end{algorithmic}
\label{alg:partial_z_correction}
\end{algorithm}

We summarize the entire decoding algorithm.
We use Algorithm~\ref{alg:direct} to compute the  syndrome for decoding.
We incorporate preprocessing techniques like peeling and clustering for better performance as in \cite{ITW_TSCC}.
After performing the preprocessing steps, we decode the bit flip errors using Algorithm~\ref{alg:dec_using_tcc_synd}, for the first stage of decoding.
In the next stage we decode the phase flip errors using Algorithm~\ref{alg:partial_z_correction}. 
The complete algorithm is given in Algorithm~\ref{alg:overall_algorithm}.
The performance of this decoder is shown in Fig.~\ref{fig:partial_gauge_fixing_result}.
We obtain a threshold of about $17.7 \%$.

\begin{remark}
For correcting the $X$ errors we fix, in effect, $3|\mathsf{F}|-6$ independent commuting operators from the gauge group
and $|\mathsf{F}|-2$ commuting operators for correcting the $Z$ errors. 
Thus a total of $4|\mathsf{F}|-8$ commuting gauge operators are fixed. 
The TSCC would have measured only $2|\mathsf{F}|-2$ operators. 
Thus by partial gauge fixing we are fixing an additional $2|\mathsf{F}|-6$  gauge operators as checks.
\end{remark}

Recall from Remark~\ref{rem:tsc-params}, that the gauge group of the TSCC is generated by $2r+s=10|\mathsf{F}|-2$ operators
where  $s=2|\mathsf{F}|-2$ is the number of independent stabilizer generators of the subsystem code and $r=4|\mathsf{F}|$ is the number of gauge qubits.
The maximal commuting subgroup in the gauge group is of size $r+s=6|\mathsf{F}|-2$.
Since we measured only $4|\mathsf{F}|-8$ checks we see that this decoder is  a partial gauge fixing algorithm. 

Observe that in this decoder, the erasure induced $X$ errors are corrected using gauge fixing while the 
$Z$ errors are not. 
This naturally suggests the design of a decoder that fixes more gauge operators. 
We take this up in the next section. 

\subsection{Correcting $Z$ errors with gauge fixing
}\label{subsec:full}
In Section~\ref{subsec:partial} we discussed an algorithm which uses partial gauge fixing to decode TSCC.
In partial gauge fixing we perform gauge fixing for clearing only bit flip errors. 
The phase flip errors are cleared by decoding the parent color code. 
The improvement in the threshold performance prompts us to fix additional gauge generators. 
We now explore the correction of $Z$ errors by gauge fixing. 
In partial gauge fixing, we fixed $4|\mathsf{F}|-8$ gauge operators out of the maximal commuting subgroup of $6|\mathsf{F}|-2$.
Hence there is a scope of fixing $2|\mathsf{F}|+6$ additional gauge operators on TSCC.
In this section we correct phase flip errors with gauge fixing, similar to bit flip error correction. 
We map the TSCC on to three copies of color codes and decode them for correcting phase flip errors.

We obtain syndromes for $X$ type stabilizers of the color codes using Algorithm~\ref{alg:direct}.
Once the color code syndromes are obtained, we decode the color codes by adapting a color code erasure decoder.
We then lift the  joint estimate on all the stacks back to the TSCC. 
We lift the error estimate back on the TSCC according to the color of the color code. 
Considering we clear both $X$ and $Z$ error on the copies of color codes, we perform gauge fixing for correcting both the errors.
We use Algorithm~\ref{alg:dec_using_tcc_synd}, with $\sigma=Z$, to correct phase flip errors with gauge fixing. 

\begin{remark}
We can see that in effect  we are fixing $6|\mathsf{F}|-12$ independent stabilizers.  
On TSCC it is possible to gauge fix  at most $r+s=6|\mathsf{F}|-2$ stabilizers.
Compared to the TSCC where we measure 
$2|\mathsf{F}|-2$ stabilizers we are measuring in addition
$4|\mathsf{F}|-10$ gauge operators. 
Thus we are order maximal with respect to the number of gauge operators that are being fixed. 
\end{remark}

From Eq.~\eqref{eqn:hypercycle-decompo}, we have that the hypercycle stabilizer of 
TSCC can be decomposed to a sum of color code stabilizers. 
Therefore, clearing the syndromes on the color codes also clears the syndrome on the 
TSCC. 
This makes sure that correction restores the state to the codespace, namely, the joint 
+1-eigenspace of the subsystem code stabilizers.
(Since color code is a CSS code, we can perform $X$ error correction and $Z$ error correction in any order.)

\subsection{Complete algorithm}
In this section we summarize both the $X$ and $Z$ correction for both approaches.
The complete algorithm is given in Algorithm~\ref{alg:overall_algorithm}.
The first step is to obtain the necessary syndromes for decoding the TSCC. 
Along with the syndromes of the color codes, we also compute the syndrome of the TSCC  which helps in the preprocessing using  peeling. 

The second step is to preprocess using peeling, see refer Algorithm~2 in \cite{ITW_TSCC} for further details on peeling. 
During peeling we correct single isolated erasures using TSCC stabilizer measurements.
Peeling results in removal of some erasures and also updates syndromes of some of the stabilizers of the TSCC and the color codes. 
Based on the location of the erasure where the correction is applied, we modify the color code syndromes as per the stack it belongs to.
An erased qubit belonging to  a face $f \in \mathsf{F}_c$ of TSCC modifies only syndromes of color code on $c$-stack. 
We also  update the hypercycle stabilizer syndromes.
Every qubit participates in three hypercycles.
Hence syndrome of these hypercycle stabilizers must be modified based on the error estimate.
For simulation simplicity we re-compute hypercycle stabilizers after every correction step during peeling.
They are used for phase flip error correction during partial gauge fixing decoding.
We do need not to keep track of the hypercycle syndromes for the order maximal gauge fixing decoder since we do not utilize them for further error correction.

After peeling, some of the erasures are corrected and the syndromes appropriately modified. 
Peeling is followed by clustering of erasures. 
We scan the lattice from left to right and top to bottom and search for a stabilizer which contains at least one erased qubit. 
We group the erased qubits of the stabilizer to form a cluster. 
(This is true even if there is a single erasure in that stabilizer.) 
For each of the erased qubit we check if it is already part of a previously formed cluster.
If so, we merge these clusters.
We repeat this procedure for every stabilizer. 
At the end, erasures of two distinct clusters do not participate in a common stabilizer.
For more details on clustering, see \cite{ITW_TSCC}. 

Once the clusters are formed, the next step is to decode every cluster independently.
For every cluster, we perform bit flip error correction using Algorithm~\ref{alg:direct}.
We then  correct the phase flip errors. 
If decoding with partial gauge fixing decoder, we use Algorithm~\ref{alg:partial_z_correction} and for order maximal gauge fixing decoder, we use Algorithm~\ref{alg:direct}.
All error estimates for each of the clusters are combined and returned at the end of the algorithm.
Note that for order maximal  gauge fixing since both bit flip error syndrome and phase flip error syndrome do not affect each other, we can decode both these errors in parallel. 

\begin{algorithm}[H]
\caption{Erasure decoder for TSCC}
\begin{algorithmic}[1]
    \REQUIRE {TSCC hypergraph $\mathcal{H}_{\Gamma}$, set of erasure positions $\mathcal{E}$ and choice of decoding: partial gauge fixing decoding or order maximal gauge fixing decoding.}
    \ENSURE {Error estimate $\widehat{E}$}
    \STATE $\widehat{E} = \mathit{I}.$
    \STATE Use Algorithm~\ref{alg:direct} to obtain syndrome $s_c^\sigma$, $\sigma \in \{X,Z\}$, $c \in \{r,g,b\}$  and syndromes of $W_1^f$ and $W_2^f$ for all $f\in \mathsf{F}$.
    \STATE Perform peeling to obtain correction operator $\widehat{E}_p$, refer \cite{ITW_TSCC}. 
    \STATE Update the color code  and TSCC stabilizers and the list of erasures. 
    \STATE Cluster remaining erasures which share a stabilizer. \COMMENT{Let the number of clusters be $n_c$} 
    \IF{$1\le i \le n_c $} 
        \STATE Use Algorithm~\ref{alg:dec_using_tcc_synd}, with $\sigma = X$, to obtain $X$ error estimate $\widehat{E}^X$.
    \IF{partial gauge fixing decoder}
    \STATE Use Algorithm~\ref{alg:partial_z_correction} to obtain $Z$ error estimate $\widehat{E}^Z$.
         \ELSE
         \STATE Use Algorithm~\ref{alg:dec_using_tcc_synd}, with $\sigma = Z$, to obtain $Z$ error estimate $\widehat{E}^Z$.
    \ENDIF
    \STATE $\widehat{E} = \widehat{E} \widehat{E}^X\widehat{E}^Z$
    \ENDIF
    \STATE Return $\widehat{E}\widehat{E}_p$ as the final error estimate.
\end{algorithmic}
\label{alg:overall_algorithm}
\end{algorithm}

We obtained a threshold of $44\%$ for the order maximal gauge fixing decoder, shown in Fig.~\ref{fig:full_gauge_fixing_result}.
Note that for both the decoding algorithms we are measuring all the color code stabilizers.
The main difference between the two algorithms is the count of gauge operators fixed.
Another crucial difference is the number of times the color code erasure decoder needs to be executed for each of the decoders.
In Table~\ref{table:diff-table} we highlight the key differences between the decoders.
We discuss the simulation results in Section~\ref{sec:simulation_results}. 
Before that, we study the correctability of  erasure patterns on TSCCs in the next section.

\begin{table*}[]
\centering
\begin{tabular}{p{2.7cm}|p{2cm}|p{5cm}|p{2.0cm}}
\hline \hline
\multicolumn{2}{c|}{\textbf{Decoder}}  & \multicolumn{1}{c|}{\textbf{Partial gauge fixing decoder}   }& \multicolumn{1}{c}{\textbf{Maximal gauge fixing decoder}                     }  \\ \hline\hline
\multicolumn{2}{l|}{Lattice for $X$ error correction}  & \multicolumn{2}{l}{Three copies of (parent) color code of the TSCC.}  
\\ \hline
\multicolumn{2}{l|}{ $X$ error correction}     & \multicolumn{2}{l}{\begin{tabular}[c]{@{}l@{}}Map erasures and syndrome to three copies of color codes. Decode them using \\ an erasure decoder.\end{tabular}}   \\ \hline
\multicolumn{2}{l|}{Lattice for $Z$ error correction}   & Parent color code of TSCC    &\begin{tabular}[c]{@{}l@{}}Three copies of (parent) color\\ code\end{tabular}   \\ \hline
\multicolumn{2}{l|}{$Z$ error correction}     & \begin{tabular}[c]{@{}l@{}}Map $W_2^f$ syndrome and erased \\ locations to the parent color code. \\ Decode using an erasure decoder\end{tabular} & \begin{tabular}[c]{@{}l@{}}Map erasures and syndrome to three \\ copies of color codes. Decode them\\  using an erasure decoder.\end{tabular} \\ \hline
\multicolumn{2}{l|}{\begin{tabular}[c]{@{}l@{}}Number of commuting gauge \\operators fixed\end{tabular}}                                           & \multicolumn{1}{c|}{ $4|\mathsf{F}|-8 $}  &  \multicolumn{1}{c}{$6|\mathsf{F}|-12  $  } \\
\hline
\multicolumn{2}{l|}
{\begin{tabular}[c]{@{}l@{}}Number of times color code\\ erasure decoder is utilized\end{tabular}}      & \multicolumn{1}{c|}{4 }     & \multicolumn{1}{c}{6}   \\ \hline
\multicolumn{2}{l|}{Errors for which gauge fixing is used}  & Only X errors  & \multicolumn{1}{l}{Both $X$ and $Z$ errors}\\ \hline
\multicolumn{2}{l|}{Threshold}      & \multicolumn{1}{c|}{ \textbf{17.7}\% }  & \multicolumn{1}{c}{ \textbf{44}\%  } \\ \hline\hline
\end{tabular}
\caption[justification=justified]{Comparison between partial gauge fixing decoder and order maximal gauge fixing decoder.}
\label{table:diff-table}
\end{table*}

\section{Correctability condition on erasure pattern}\label{sec:bound}
In this section we propose  correctabilty conditions for an erasure pattern on subsystem codes. 
Using the correctability condition we can test, prior to decoding, whether an erasure pattern can be corrected or not. 
In Theorem~\ref{th:correct-condition}, we propose the condition for a general subsystem code. 

Before we begin, we introduce a few notations necessary for this section. 
We denote the stabilizer matrix of the subsystem code by $H$ and matrix representation of the gauge group over $\mathbb{F}_2$ by $G$.
The stabilizer matrix $H$ involves all the (independent) stabilizer generators of the subsystem code 
while $G$ consists of the gauge generators.
Let $H_{\mathcal{E}}$ be a submatrix which is a restriction of $H$ to the subset of  columns corresponding to locations of erased qubits. 
Similarly, we denote by $G_\mathcal{E}$ and $\mathcal{E}_{\bar{\mathcal{E}}}$, the submatrices of $G$
restricted to the qubits in $\mathcal{E}$ and $\bar{\mathcal{E}}$ respectively.
The submatrix $H_{\bar{\mathcal{E}}}$ consists of the remaining columns of $H$.
We denote the stabilizer matrix of the parent color code by $H_c$, where $c \in \{r,g,b\}$. 

An erasure pattern $\mathcal{E}$ is correctable if all the errors supported in $\mathcal{E}$ are correctable and non-correctable otherwise. 
Equivalently, if $\mathcal{E}$ supports a logical error, then it is not correctable and correctable otherwise. 
A necessary condition for an erasure pattern  to be correctable for stabilizer codes due to 
\cite{delfosse2012upper} is as follows:
\begin{proposition} [\cite{delfosse2012upper}]\label{prop:delfosse-condition}
For stabilizer codes, an erasure pattern $\mathcal{E}$ is correctable only if
\begin{eqnarray}
2|\mathcal{E}| \leq \text{rank}(H) + \text{rank}(H_{\mathcal{E}})-\text{rank}(H_{\bar{\mathcal{E}}}) \label{eq:CCcorrectableE}
\end{eqnarray}
\end{proposition}

 {\em Proposition~\ref{prop:delfosse-condition} immediately gives a sufficient condition for a correctable erasure pattern decoded via order maximal gauge fixing decoder.} 
Specifically, if Eq.~\eqref{eq:CCcorrectableE} is  satisfied on all the three copies of color code i.e.,
\begin{eqnarray}
2|\mathcal{E}_c| \leq \text{rank}(H_c) + \text{rank}(H_{\mathcal{E}_c})-\text{rank}(H_{\bar{\mathcal{E}_c}}) \label{eq:CCcorrectableE_order_maximal}
\end{eqnarray}
for every $c$, where $c \in \{r,g,b\}$, then the erasure is correctable by the order maximal gauge fixing decoder. 
Note  that this is only  a  sufficient  condition for the correctability of an erasure pattern and not a necessary condition, see the  discussion in Section~\ref{subsec:residual-errors}.

We now propose a condition for correctable erasures applicable to general subsystem codes decoded without gauge fixing. 
This condition also accounts for the gauge group.
\begin{theorem}[Correctable erasures on a subsystem code without gauge fixing] \label{th:correct-condition}
An erasure pattern $\mathcal{E}$ on a TSCC is correctable only if 
the following condition is satisfied. 
\begin{eqnarray}
2|\mathcal{E}| = \text{rank}(H_{\mathcal{E}})+\text{rank}(G)-\text{rank}(G_{\bar{\mathcal{E}}}) \label{eq:correctableE}
\end{eqnarray}
\end{theorem}
\begin{proof}
Up to a phase, there are $4^{|\mathcal{E}|}$ Pauli errors that can be supported in $\mathcal{E}$.
There are $2^{\text{rank}(H_{\mathcal{E}})}$ syndromes possible. 
For each syndrome there exist $2^{2|\mathcal{E}|-\text{rank}(H_{\mathcal{E}})}$ error patterns. 
Consider the map 
\begin{eqnarray}
\pi_{\bar{\mathcal{E}}}:G\rightarrow G_{\bar{\mathcal{E}}}
\end{eqnarray}
The operators in the kernel of $\pi_{\bar{\mathcal{E}}}$ are exactly the operators in $G$ whose support is entirely in $\mathcal{E}$.
The dimension of $\ker{\pi}_{\bar{\mathcal{E}}}$ is given by
\begin{eqnarray}
\ker(\pi_{\bar{\mathcal{E}}})=\text{rank} (G)-\text{rank}(G_{\bar{\mathcal{E}}})
\end{eqnarray}
All errors in $\ker(\pi_{\bar{\mathcal{E}}})$ have support only in $\mathcal{E}$
and have zero syndrome. 
First observe that such errors cannot be more than the total number of errors supported in 
$\mathcal{E}$ which have zero syndrome. 
Therefore, $\text{rank} (G)-\text{rank}(G_{\bar{\mathcal{E}}}) \leq 2|\mathcal{E}| -\text{rank}(H_\mathcal{E})$.
If errors in $\ker(\pi_{\bar{\mathcal{E}}})$ are fewer than the number of distinct errors with the same syndrome, then $\mathcal{E}$ supports  nontrivial logical error(s). 
Therefore, if $\text{rank} (G)-\text{rank}(G_{\bar{\mathcal{E}}}) < 2|\mathcal{E}| -\text{rank}(H_\mathcal{E})$, then $\mathcal{E}$ is not correctable.
If $2|\mathcal{E}| -\text{rank}(H_\mathcal{E}) = \text{rank} (G)-\text{rank}(G_{\bar{\mathcal{E}}})$, 
all such errors are in the gauge group.
So $\mathcal{E}$ does not support a (nontrivial) logical operator and  $\mathcal{E}$ is a correctable erasure pattern.
\end{proof}

\begin{remark}
Since it is not possible for $\text{rank} (G)-\text{rank}(G_{\bar{\mathcal{E}}}) > 2|\mathcal{E}| -\text{rank}(H_\mathcal{E})$,
we can state the correctability condition in Theorem~\ref{th:correct-condition} slightly differently. 
More precisely, an erasure is correctable only if 
\begin{eqnarray}
2|\mathcal{E}| \leq  \text{rank}(H_{\mathcal{E}})+\text{rank}(G)-\text{rank}(G_{\bar{\mathcal{E}}}) \label{eq:correctableE-alt}
\end{eqnarray}
In this form it reduces to the condition in \cite[Eq.~(4)]{delfosse2012upper}, when 
$G$ is Abelian i.e., we have a stabilizer code.
\end{remark}

\begin{figure}[ht]
    \centering
\includegraphics[scale=1]{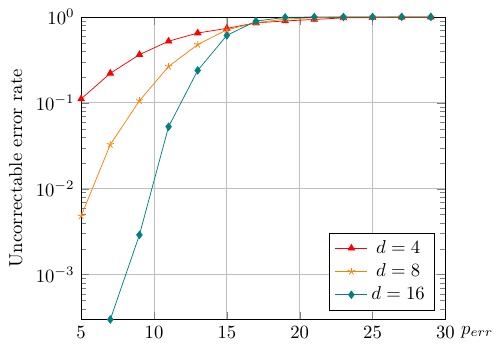}
    \caption{Threshold of $\approx 16.5\%$ based on Eq.~\eqref{eq:correctableE}.}
    \label{fig:correctability_condition_results}
\end{figure}

We performed simulations based on the condition in Eq.~\eqref{eq:correctableE}. 
We generate an erasure pattern $\mathcal{E}$ where qubit is erased with a probability $\varepsilon$ on the subsystem code. 
As per the locations of  $\mathcal{E}$, we compute the rank of the matrices $H_{\mathcal{E}}$ and $G_{\bar{\mathcal{E}}}$. 
Rank of $G$ remains constant throughout the simulations since it is independent of $\mathcal{E}$. 
If the condition in Eq.~\eqref{eq:correctableE} gets violated, we flag it as an uncorrectable error. 
We repeated the experiment for 10000 trials for various $\varepsilon$ for code distances $d={4,8,16}$.
We obtained a threshold of approximately $16.5\%$ as shown in Fig.\ref{fig:correctability_condition_results}.

\section{Simulation results}\label{sec:simulation_results}
In this section we present the simulation results for both the decoding algorithms. 
We simulate for the  TSCC derived from square octagon lattice.
The proposed algorithms can be easily adapted to other TSCCs.

\subsection{Simulation setup}
An erasure pattern is generated on the subsystem codes according to the 
probability of erasure $\varepsilon$. 
Noise on each qubit is assumed to be identical and independent. 
Then using the erasure model described in Section~\ref{subsec:errormodel}, we generate Pauli errors on each of the erased qubits with equal probability. 
Every erased location can undergo any of the Pauli errors with equal probability as described in Section~\ref{subsec:errormodel}.
Then we correct these induced Pauli errors on the subsystem code.
We use Algorithm~\ref{alg:overall_algorithm} for decoding.

In practice, while decoding a subsystem code,  its stabilizers are measured indirectly via measuring the gauge generators and classically combining the outcomes appropriately. 
The proposed decoders  map the TSCCs to multiple copies of color codes in order to decode them.
We measure the multi-body color code stabilizers \textit{directly} on the TSCC.
Once the color code stabilizers are measured, we decode the three copies of color code using the color code erasure decoder \cite{aloshious2019erasure}.
This decoder only needs the erasure locations as input and generates its own syndrome. 
We adapted it so that takes as input the previously measured syndrome and the erasure locations.

After decoding the color code copies, we lift the estimate to TSCC.
After lifting we check if any logical error has occurred or not. 
If the product of original error and the error estimate anticommutes with any one of the bare logical operators, we declare a logical error.

\subsection{Results}
\begin{figure}[H]
    \centering
\includegraphics[scale=1]{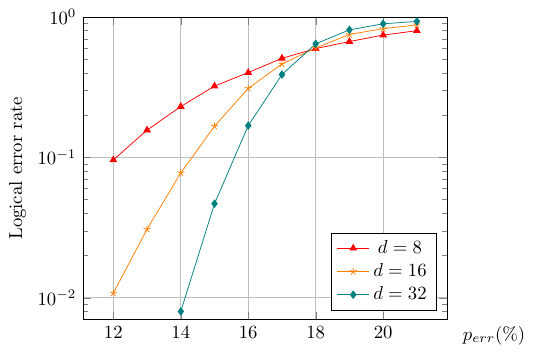}
    \caption{Performance of partial gauge fixing decoder. Threshold is $\approx 17.7\% $.}
    \label{fig:partial_gauge_fixing_result}
\end{figure}

\begin{figure}[H]
    \centering
\includegraphics[scale=1]{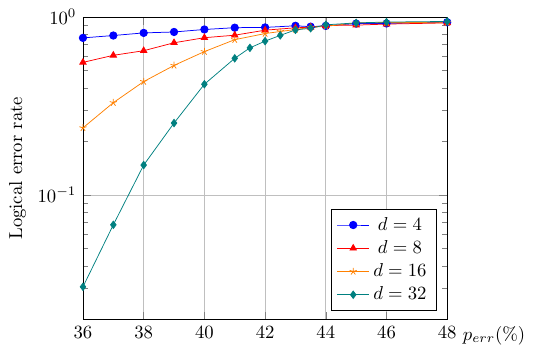}
    \caption{Performance of the order maximal gauge fixing decoder. Threshold is $\approx 44\% $.}
    \label{fig:full_gauge_fixing_result}
\end{figure}

We present the simulation results for the  TSCC derived from square octagon lattice.
The TSCCs derived from the square octagon lattices have the code parameters $[[3d^2,2,2d^2,d]]$, where $d$ is the distance of the code \cite{bombin2010topological,ps2012}. 
We have simulated the proposed decoders for these codes for  various erasure probabilities and
lattice sizes. 
The gauge measurements are assumed to be noiseless. 
The plots shown in Fig.~\ref{fig:partial_gauge_fixing_result} and Fig.~\ref{fig:full_gauge_fixing_result} show the variation of logical error rate with respect to the probability of erasure errors. 
A logical error occurs when the error and error estimate differ by a nontrivial logical operator
of the TSCC. 
Every data point has been obtained by accumulating $2000$ logical errors or $10000$ runs, whichever occurs earlier.

Fig.~\ref{fig:partial_gauge_fixing_result} shows the plot for performance of the partial gauge fixing decoder. 
We report a threshold of approximately $17.7\%$.
This suggests that gauge fixing improves the performance in comparison to the highest threshold of $9.7\%$ given in \cite{ITW_TSCC}.
When extending to almost full gauge fixing, we observe a threshold of $44\%$ as shown in Fig.~\ref{fig:full_gauge_fixing_result}.
The performance of our decoder is close to $50\%$ which is set by the no-cloning theorem.

\section{Conclusion}\label{sec:conclusion}
In this paper, we proposed two algorithms for decoding subsystem codes over erasure channel.
By using the technique of gauge fixing in combination with other preprocessing techniques we were able to 
significantly improve the threshold of TSCCs for the erasure noise with respect to our
previous work \cite{ITW_TSCC}.
Our work draws upon the mapping of TSCC to multiple copies of color codes proposed in \cite{bombin2012universal}.
We decode on these color codes instead of  decoding directly on TSCC
which motivates in a sense the need for gauge fixing. 
Our first decoder uses partial gauge fixing where gauge fixing is used only to correct erasure induced bit flip errors.
The second decoder uses maximal gauge fixing where both bit flip and phase flip errors are corrected via gauge fixing.
The later decoder gives us a threshold of $44\%$. 
To the best of our knowledge, this is the highest threshold to date for a TSSC for erasure noise.
Syndrome measurement for both the decoders requires us to measure 4-body and 8-body measurements.
We also formulated conditions for correctability of erasures on the subsystem codes. 
There remain many other interesting open questions in this context.
Developing an optimal decoder without gauge fixing, 
improving the performance of the proposed decoders, reducing the complexity of syndrome measurement
are some natural problems for further study. 

\medskip
\noindent
{\em Acknowledgment.}
HMS would like to thank Arun B. Aloshious for valuable discussions. 


%

\end{document}